\documentclass[a4paper,11pt]{article}
\usepackage{pos}
\usepackage{lineno}
\usepackage{hyperref}
\usepackage{caption}
\usepackage{subcaption}
\usepackage{wrapfig}

\title{System-size dependence of particle production at mid- and forward rapidity with ALICE}

\manuallySeparateAuthors

\author*[a]{Abhi Modak}
\author{ for the ALICE Collaboration}

\affiliation[a]{Bose Institute,\\
  EN 80, Sector V, Bidhan Nagar, Kolkata, India}

\emailAdd{abmodak@cern.ch}

\abstract{The pseudorapidity densities of charged particles and inclusive photons produced
  in high energy nuclear collisions are essential observables to characterise the global
  properties of the collisions, such as the achieved energy density, and to provide important
  constraints for Monte Carlo model calculations.
  In the LHC Run~1 and Run~2 configurations, ALICE had large coverage to measure charged
  particles over a pseudorapidity range ($-3.4~<~\eta~<~5.0$), combining the data from
  the Silicon Pixel Detector (SPD) and the Forward Multiplicity Detector (FMD). The inclusive
  photons are measured at forward rapidity using the Photon Multiplicity Detector (PMD),
  covering the pseudorapidity range $2.3~<~\eta~<~3.9$. New results on charged-particle
  pseudorapidity densities measured in pp, p--Pb, and Pb--Pb collisions at
  $\sqrt{s_{\rm NN}}$~=~5.02~TeV using Run~1 and Run~2 data are presented. Inclusive photon
  production is reported for p--Pb collisions at $\sqrt{s_{\rm NN}}$~=~5.02~TeV.
  The charged-particle rapidity densities are derived from the measured charged-particle
  pseudorapidity densities, and then parameterized by a normal distribution. This allows us
  to study the evolution of the width of the rapidity distributions as a function of the
  number of participants in all three collision systems. The performance of the new Inner Tracking
  System (ITS) designed for ALICE Run~3 configuration is also discussed for pilot beam pp collisions
  at $\sqrt{s}$ = 0.9~TeV.}

\FullConference{%
  41st International Conference on High Energy physics - ICHEP2022\\
  6-13 July, 2022\\
  Bologna, Italy
}

\begin{document}
\maketitle
\section{Introduction}
\vspace{-0.2cm}
Particle production at the Large Hadron Collider (LHC) energies is driven by a combination
of hard (perturbative) and soft (non-perturbative) quantum chromodynamics (QCD) processes.
Soft QCD processes dominate the bulk of particle production at low transverse momenta and
can only be described by phenomenological models and effective theories. Multiplicity and
pseudorapidity distributions of the produced final-state particles are some of the basic
measurements to characterise the global properties of the collisions~\cite{Bjorken:1982qr}
and to provide constraints for better tuning of models in understanding the underlying
description of particle production. During the LHC Run~1 and Run~2, ALICE recorded data with
various colliding systems (pp, p--Pb, Xe--Xe and Pb--Pb) at different center-of-mass energies.
This offered the possibility to study the evolution of particle production with collision
energy and system-size which will further help to learn about how the nuclear medium affects
particle production mechanisms.

In this article, we report measurements of charged-particle pseudorapidity
densities ($\mathrm{d}N_\mathrm{ch}/\mathrm{d}\eta$) in pp, p--Pb, and Pb--Pb collisions at
mid- and forward rapidities using the ALICE detector~\cite{ALICEDet}. The inclusive photon
production, which provides the information complementary to those obtained from the charged
particles, is studied in p--Pb collisions over the kinematic range $2.3~<~\eta~<~3.9$.
The performance of the new Monolithic Active Pixel Sensors-based Inner Tracking System (ITS)
(designed for ALICE Run~3 configuration~\cite{ALICEDetUpgrade}) and the tracking/matching
algorithms are presented for LHC pilot run of pp collisions at $\sqrt{s}$ = 0.9~TeV in
October 2021.

\section{Analysis method}
\vspace{-0.2cm}
The measurements of $\mathrm{d}N_\mathrm{ch}/\mathrm{d}\eta$ at $\sqrt{s_{\rm NN}}$~=~5.02~TeV were
performed based on the data collected by ALICE during LHC Run~1 for p--Pb collisions (in 2013)
and during Run~2 for pp and Pb--Pb collisions (in 2015). The p--Pb and Pb--Pb data were analysed with a
minimum bias (MB) trigger requiring a coincidence of signals in each side of the V0 sub-detectors
(V0A and V0C). Likewise, the pp data were analysed for inelastic (INEL) interactions with at
least one charged-particle detected at $|\eta|~<~1$ (INEL~$>$~0 event class). The standard ALICE
event selection and centrality determination based on the V0 amplitude were considered in this
analysis~\cite{ChPrSysSize,ALICECentSelect}. The $\mathrm{d}N_\mathrm{ch}/\mathrm{d}\eta$ was
measured by counting the number of tracklets using the SPD detector at mid-rapidity ($|\eta|~<~2$).
At forward rapidity the measurement was done based on the deposited energy signal in the FMD and
a statistical method was employed to calculate the inclusive number of charged
particles~\cite{ChPrSysSize}. The detailed studies on the estimation of systematic uncertainties
in pp, p--Pb and Pb--Pb collisions can be found in Ref.~\cite{ChPrSysSize}.

The $\mathrm{d}N_\mathrm{ch}/\mathrm{d}\eta$ was also analysed, using the new ALICE computing software
framework, named Online-Offline (O$^2$)~\cite{ALICEo2}, for pp collisions at $\sqrt{s}$ = 0.9~TeV using
the pilot beam Run~3 data collected in October 2021. The events were selected using timing
information of the Fast Interaction Trigger (FIT) detector~\cite{ALICEFit}. The charged-particle
tracks that have at least one hit in any of the ITS layers were considered in this analysis.
Correction procedures similar to the ones described in~\cite{ALICEppChPaper} are used to
correct the measured distributions. Systematic uncertainties from various sources (event generator
dependence, $p_{\rm T}$ uncertainties, strangeness corrections and particle composition)
were evaluated using techniques identical to those reported in~\cite{ALICEppChPaper}.

The pseudorapidity density of inclusive photons ($\mathrm{d}N_\mathrm{\gamma}/\mathrm{d}\eta$) was
measured on an event-by-event basis using a preshower technique with the PMD
detector~\cite{PMDNimpaper,PMDppPaper,PMDpPb}. The reconstruction of photons consists of two steps:
(i) finding clusters of hits on the preshower plane of the PMD and (ii) discriminating between
photons and charged hadrons. The clustering was performed using a nearest neighbour clustering
algorithm. Each cluster is characterized by the number of cells (cluster $N_{\rm cell}$) contained
in it and the total energy (measured in terms of ADC) deposited in that cluster. Suitable
photon-hadron discrimination thresholds were applied based on the cluster $N_{\rm cell}$
and ADC content to obtain a photon-rich sample, known as $\gamma$-like
clusters~\cite{PMDppPaper}. A Bayesian unfolding technique~\cite{BayesUnfold}
was then used to correct the measured distributions of $\gamma$-like clusters affected by
instrumental effects and the contamination from hadron clusters~\cite{PMDpPb}. A detailed
discussion on the estimation of systematic uncertainties for the photon analysis
is given in Refs.~\cite{PMDppPaper,PMDpPb}.

\section{Results and discussion}
\begin{wrapfigure}{r}{0.6\textwidth}
  \begin{center}
    \includegraphics[scale=0.55]{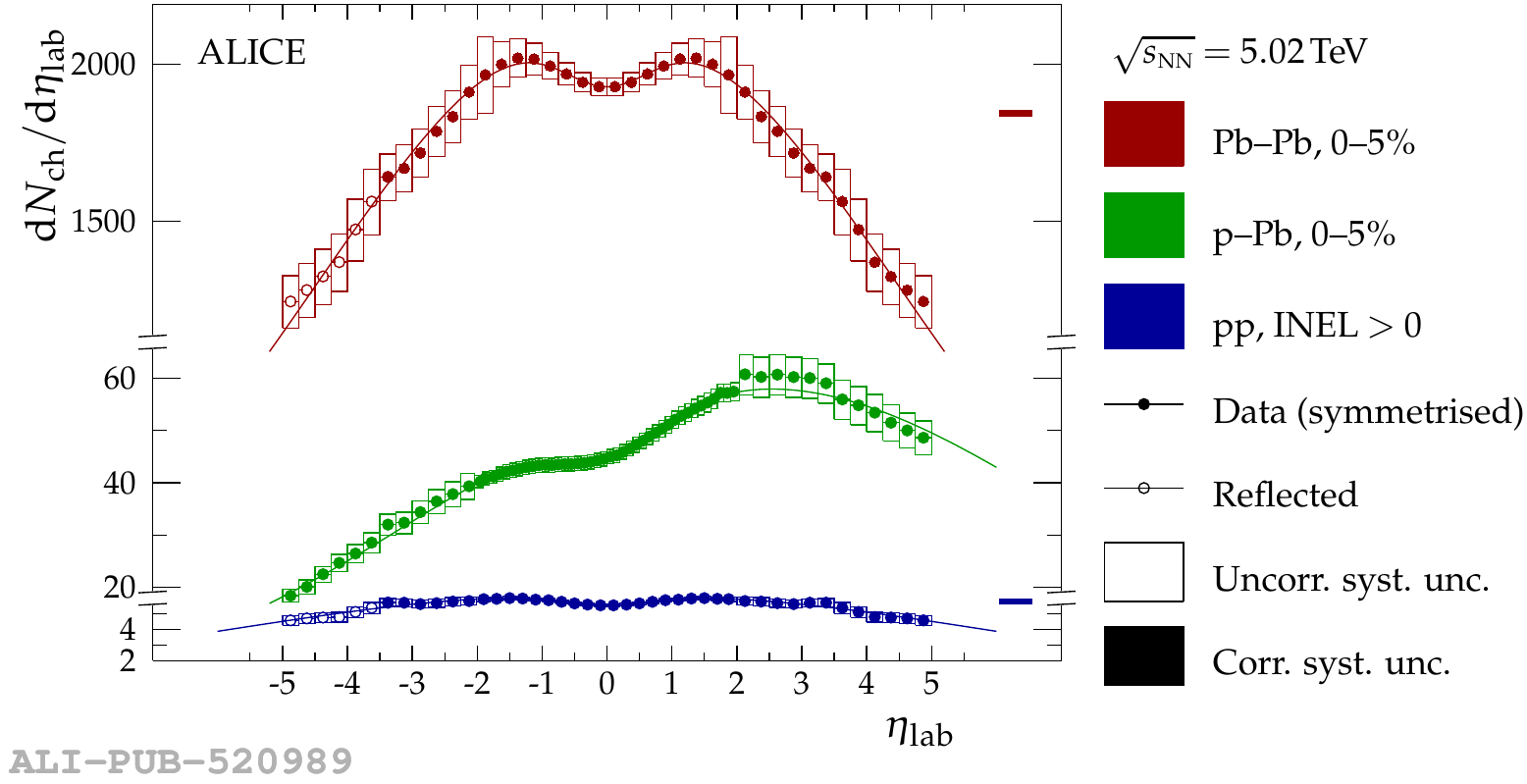}
  \end{center}
  \vspace*{-0.5cm}
  \caption{Charged-particle pseudorapidity density for minimum bias (INEL~$>$~0) pp and
    for the 5\% most central p--Pb and Pb--Pb collisions at $\sqrt{s_{\rm NN}}$~=~5.02~TeV.
    The lines show fits of Eq.~\ref{fit_pp_PbPb} (Pb--Pb and pp) and Eq.~\ref{fit_p_Pb} (p--Pb)
    to the data (see text). Please note that the ordinate has been cut twice to accommodate
    for the very different ranges of the $\mathrm{d}N_\mathrm{ch}/\mathrm{d}\eta$~\cite{ChPrSysSize}.}
  \label{dnchdeta_all}
\end{wrapfigure}

\vspace{-0.2cm}
Figure~\ref{dnchdeta_all} shows the primary charged-particle pseudorapidity
densities measured in pp collisions with the INEL~$>$~0 event class and in
top central (0--5\%) p--Pb and Pb--Pb collisions at $\sqrt{s_{\rm NN}}$~=~5.02~TeV.
The $\mathrm{d}N_\mathrm{ch}/\mathrm{d}\eta$ at mid-rapidity in pp collisions is
found to be 5.7 $\pm$ 0.2 while for 0--5\% Pb--Pb collisions it reaches
$\approx$ 2000. A clear asymmetric shape of $\mathrm{d}N_\mathrm{ch}/\mathrm{d}\eta$
is observed in p--Pb collisions and the distribution peaks at
$\mathrm{d}N_\mathrm{ch}/\mathrm{d}\eta$~$\approx$~60~around~$\eta$~=~3 on the Pb-going
direction.

\begin{figure}[h!]
  \centering
  \begin{subfigure}[b]{0.4\textwidth}
    \centering
    \includegraphics[scale=0.4]{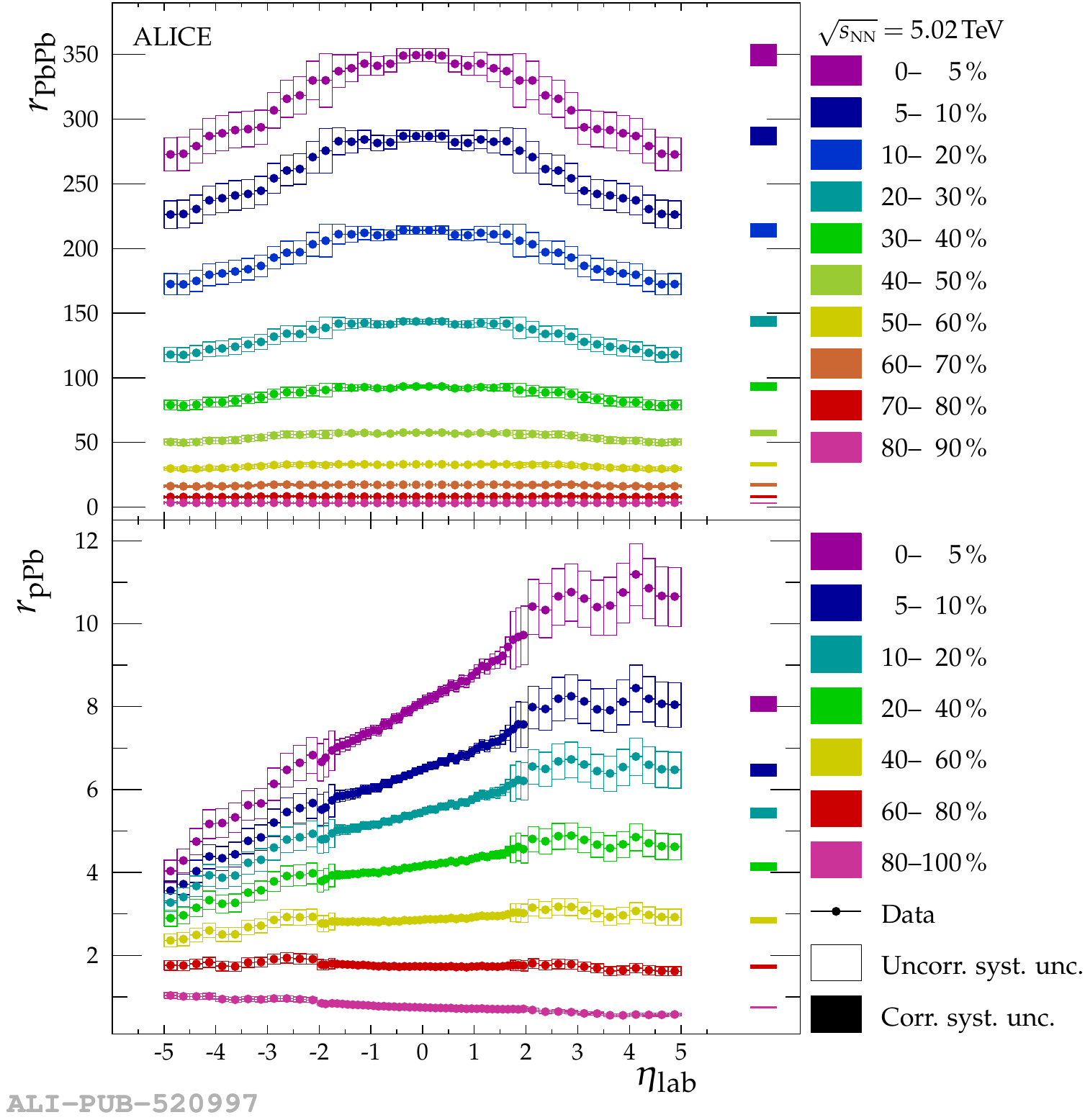}
    \caption{}
    \label{dnchdeta_ratio}
  \end{subfigure}
  \hspace*{1.cm}
  \begin{subfigure}[b]{0.4\textwidth}
    \centering
    \includegraphics[scale=0.4]{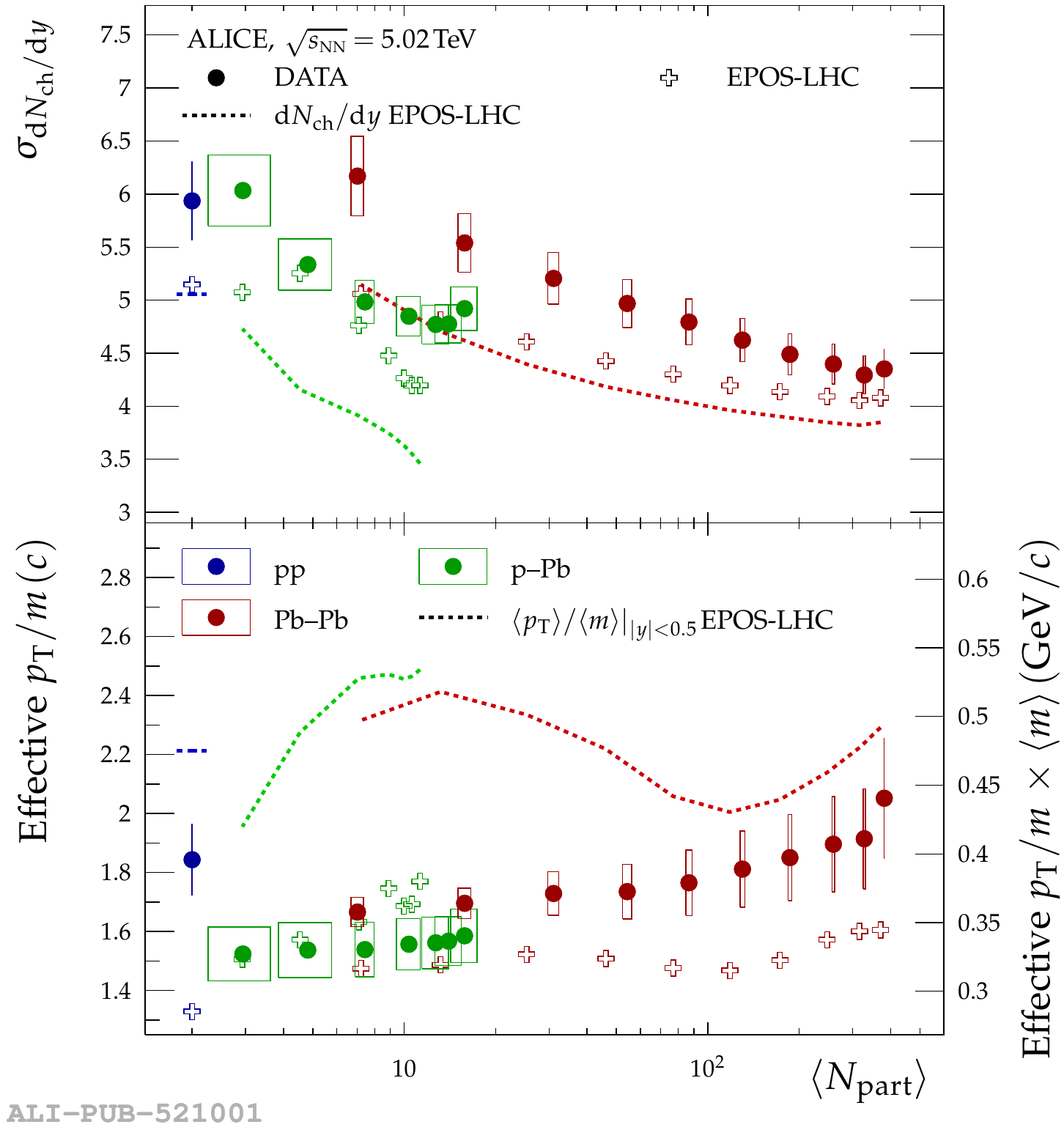}
    \caption{}
    \label{dnchdeta_width}
  \end{subfigure}
  \caption{(a) The ratio of the charged-particle pseudorapidity density measured in Pb--Pb (top) and
    p--Pb (bottom) in different centrality classes to the same quantity obtained in pp for the INEL~$>$~0
    event class. Note, for Pb--Pb $\eta_{\rm lab}$ is the same as the center-of-mass pseudorapidity.
    (b) The width (top) and effective $p_{\rm T}/m$ (bottom) extracted from best-fit parameterizations
    using Eq.~\ref{fit_pp_PbPb} and Eq.~\ref{fit_p_Pb} are shown as a function of mean number of
    participants. Similar fit parameters from the same parameterisation of EPOS-LHC calculations
    are also shown~\cite{ChPrSysSize}.}
\end{figure}

The ratio of the charged-particle pseudorapidity density,
$r_{X} = (\mathrm{d}N_\mathrm{ch}/\mathrm{d}\eta|_{X})/(\mathrm{d}N_\mathrm{ch}/\mathrm{d}\eta)|_{\rm pp}$
(where $X$ labels centrality classes in p--Pb and Pb--Pb collisions), in p--Pb and Pb--Pb collisions
to that in pp collisions is presented as a function of $\eta$ in Fig.~\ref{dnchdeta_ratio}.
It is observed that the ratio, $r_{\rm pPb}$, increases linearly with $\eta$ from the p-going
to the Pb-going direction for central collisions which suggests a scaling of the pp distribution
with the increasing number of participants as the lead nucleus is probed by the incident proton.
A similar scaling, however, is not observed in the ratio of the Pb--Pb distributions relative
to the pp. The $r_{\rm PbPb}$ exhibits an enhancement of particle production around $\eta$ = 0
for the more central collisions.

The charged-particle rapidity density ($\mathrm{d}N_\mathrm{ch}/\mathrm{d}y$) in Pb--Pb
collisions~\cite{ALICE:2016fbt}, to a good accuracy, is observed to follow a normal distribution
within $|y|~\lesssim~5$. The conversion of $\mathrm{d}N_\mathrm{ch}/\mathrm{d}\eta$ to
$\mathrm{d}N_\mathrm{ch}/\mathrm{d}y$ is given by
$\mathrm{d}N_\mathrm{ch}/\mathrm{d}\eta = \bigl(1 + \frac{m^2}{p_{\rm T}^2 \cosh^2\eta}\bigr)^{-1/2}
\mathrm{d}N_\mathrm{ch}/\mathrm{d}y$.
Hence, $\mathrm{d}N_\mathrm{ch}/\mathrm{d}\eta$ for symmetric collision systems (pp and Pb--Pb)
can be parameterized as~\cite{ChPrSysSize,ALICE:2016fbt}
\begin{equation}
  f(\eta; A, a, \sigma) = \bigl(1 + 1/a^2 \cosh^2\eta\bigr)^{-1/2} \bigl(A/\sqrt{2\pi}\sigma\bigr)
  \exp\bigl(- y^2(\eta, a)/2\sigma^2\bigr)
  \label{fit_pp_PbPb}
\end{equation}
where $a = p_{\rm T}/m$, $A$ and $\sigma$ are the total integral and width of the distribution,
respectively, and $y$ the rapidity in the center-of-mass frame. Based on the observation of
the linear increase of the p--Pb to pp ratios, the $\mathrm{d}N_\mathrm{ch}/\mathrm{d}\eta$ for
the asymmetric system (p--Pb) can also be parameterized as
\begin{equation}
  g(\eta; A, a, \beta, \sigma) = \bigl(1 + 1/a^2 \cosh^2\eta\bigr)^{-1/2} \bigl[(\beta y(\eta, a) + A)/\sqrt{2\pi}\sigma\bigr]
  \exp\bigl(-[y(\eta, a)-y_{\rm CM}]^2/2\sigma^2\bigr)
  \label{fit_p_Pb}
\end{equation}
where $A$ is replaced by $(\beta y(\eta, a) + A)$ which is linear in $y$ multiplied by a factor $\beta$.
and $y_{\rm CM} = 0.465$ which is the rapidity of the center-of-mass system in the laboratory frame.
Best-fit parameterizations of the measured $\mathrm{d}N_\mathrm{ch}/\mathrm{d}\eta$ in terms of
Eq.~\ref{fit_pp_PbPb} (for pp and Pb--Pb) and Eq.~\ref{fit_p_Pb} (for p--Pb) are shown in
Fig.~\ref{dnchdeta_all}. The charged-particle pseudorapidity distributions are well described
by these two functions $f$ and $g$, indicating that the particle production in pp, p--Pb, and
Pb--Pb collisions follow a normal distribution in rapidity.

The top panel of Fig.~\ref{dnchdeta_width} presents the width of the charged-particle rapidity
distributions ($\sigma_{\mathrm{d}N_\mathrm{ch}/\mathrm{d}y}$) for pp, p--Pb, and Pb--Pb collisions as a
function of the average number of participating nucleons ($\langle N_{\rm part} \rangle$)
calculated using a Glauber model. The $\sigma_{\mathrm{d}N_\mathrm{ch}/\mathrm{d}y}$ extracted from 
the same parameterisation of the EPOS-LHC calculations~\cite{eposlhc} is shown by open markers.
The dashed lines are obtained directly by evaluating the width of the charged particle
rapidity density from EPOS-LHC model. The general trend is that the
$\sigma_{\mathrm{d}N_\mathrm{ch}/\mathrm{d}y}$ decreases as $\langle N_{\rm part} \rangle$ increases,
consistent with the behaviour of the Pb--Pb to pp ratios. It is also noted that the width of the
$\mathrm{d}N_\mathrm{ch}/\mathrm{d}y$ distributions in p--Pb and Pb--Pb approaches that of the pp
distribution at low $\langle N_{\rm part} \rangle$. These results suggest that the enhancement of
particle production near mid-rapidity in Pb--Pb is an effect of the nuclear medium. The bottom
panel of Fig.~\ref{dnchdeta_width} shows the dependence of the parameter $a$ on
$\langle N_{\rm part} \rangle$. The right-hand ordinate is the same but multiplied by the average
$\langle m \rangle = (0.215 \pm 0.001)~\rm{GeV}/c^2$. The parameter $a$ extracted from the
EPOS-LHC calculations~\cite{eposlhc} is also presented (open markers) in the figure. The dashed
lines represent the average $p_{\rm T}/m$ predicted by the EPOS-LHC~\cite{eposlhc}. The model
calculations indicate that the extracted transverse momentum to mass ratio $a$ is smaller than the
$\langle p_{\rm T}\rangle/\langle m \rangle$.

\begin{figure}[h!]
  \hspace*{-1cm}
  \centering
  \begin{subfigure}[b]{0.4\textwidth}
    \centering
    \includegraphics[scale=0.35]{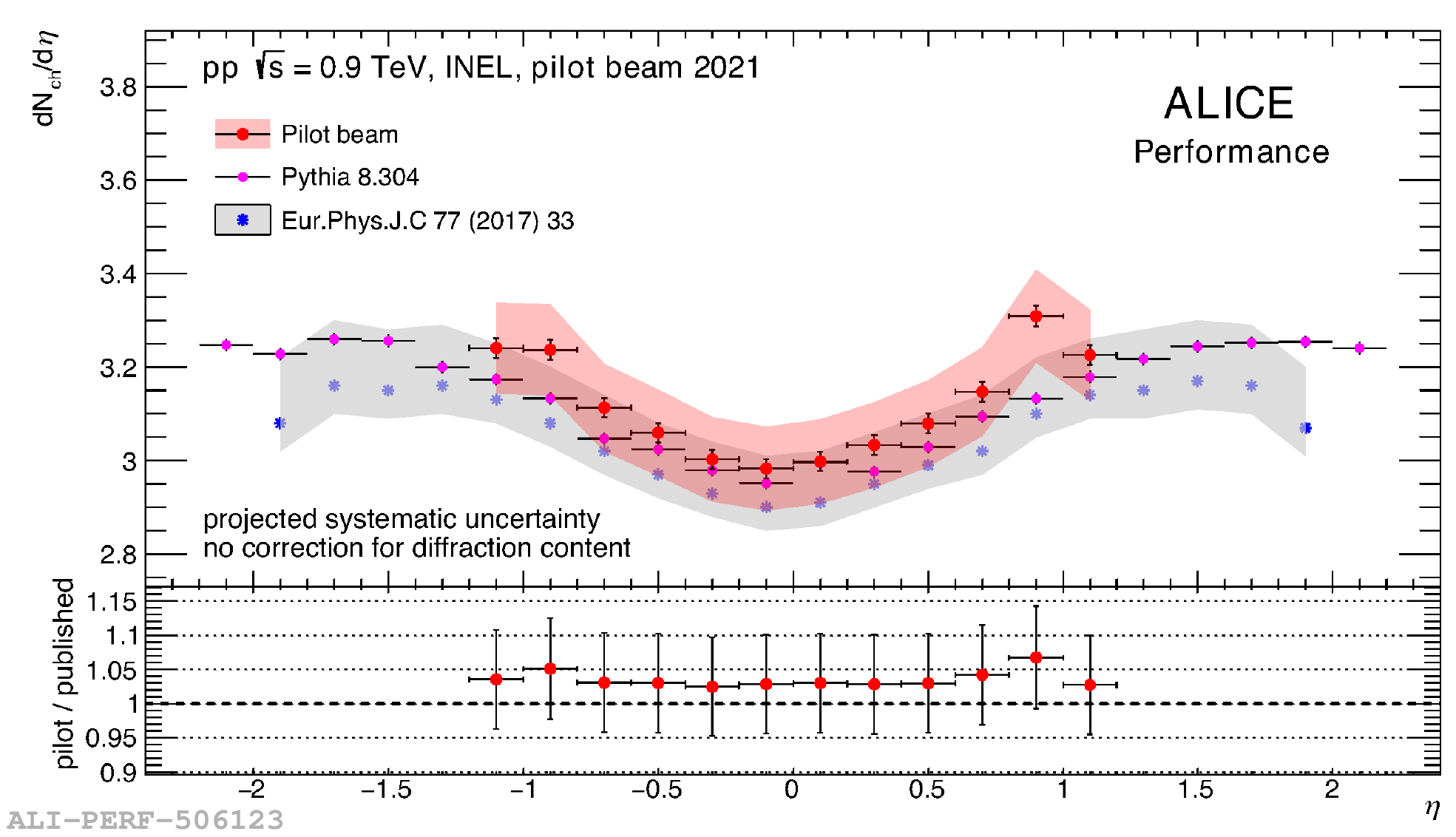}
    \caption{}
    \label{dnchdeta_pp_inel}
  \end{subfigure}
  \hspace*{1.cm}
  \begin{subfigure}[b]{0.4\textwidth}
    \centering
    \includegraphics[scale=0.35]{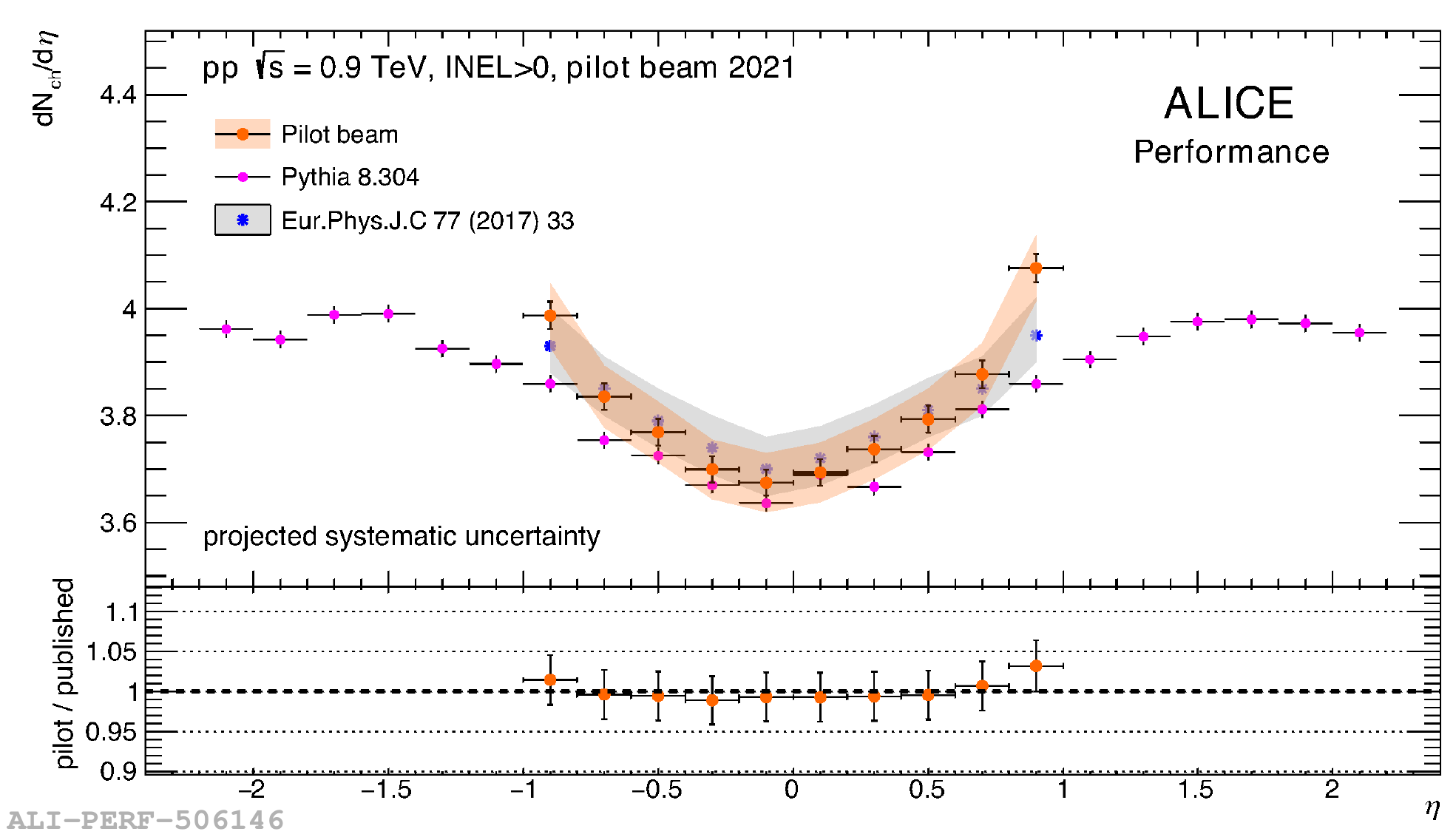}
    \caption{}
    \label{dnchdeta_pp_inelgt0}
  \end{subfigure}
  \caption{Charged-particle pseudorapidity densities in pp collisions at $\sqrt{s}$ = 0.9~TeV
    for INEL (a) and INEL~$>$~0 (b) event classes are presented together with ALICE's previous
    measurements~\cite{ALICE:2015olq} and PYTHIA~\cite{pythia} model predictions.}
  \label{dnchdeta_pp}
\end{figure}

\begin{wrapfigure}{r}{0.5\textwidth}
  \vspace*{-0.5cm}
  \begin{center}
    \includegraphics[scale=0.33]{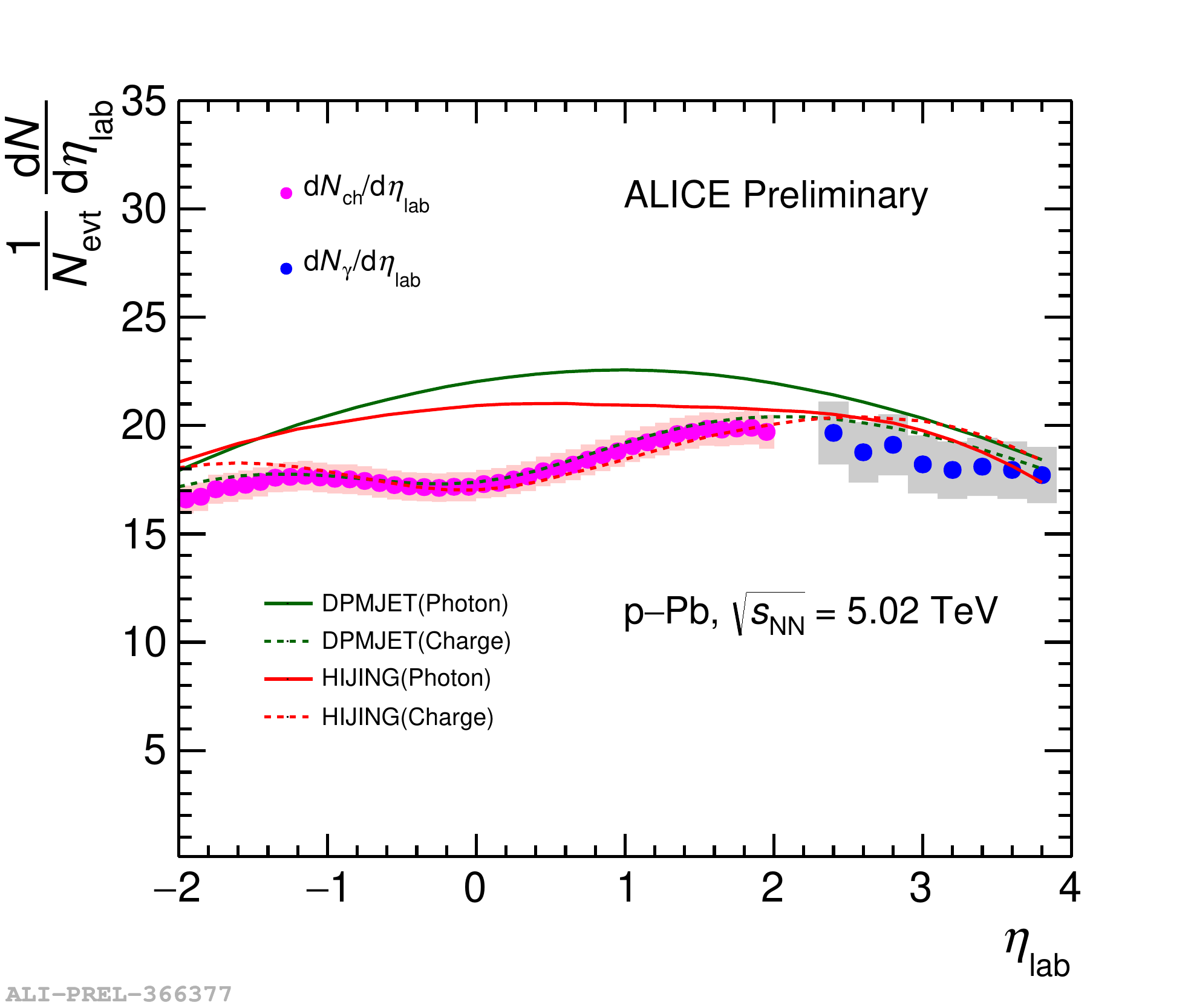}
  \end{center}
  \vspace*{-0.5cm}
  \caption{$\mathrm{d}N_\mathrm{\gamma}/\mathrm{d}\eta$ measured in p--Pb collisions at
    $\sqrt{s_{\rm NN}}$~=~5.02~TeV for MB events. The results are compared with similar
    measurements for charged particles at mid-rapidity~\cite{ChPrpPb}. Predictions
    from HIJING~\cite{HIJING} and DPMJET~\cite{DPMJET} are superimposed.}
  \label{dngammadeta_mb}
\end{wrapfigure}

Figures~\ref{dnchdeta_pp_inel} and~\ref{dnchdeta_pp_inelgt0} present the measurements of
$\mathrm{d}N_\mathrm{ch}/\mathrm{d}\eta$ using Run~3 pilot beam data in pp collisions
at $\sqrt{s}$ = 0.9~TeV for INEL and INEL~$>$~0 event classes respectively. The results
are compared to the published Run~2 measurements~\cite{ALICE:2015olq}. The difference between
Run~2 and Run~3 measurements comes mainly from: (a) the upgraded ALICE detector and (b) new
reconstruction algorithms. It is observed that there is good agreement
between the present measurements and published results.
However, the measurements for the INEL event class is slightly higher than the published data
due to the lack of diffraction correction in the Run~3 Monte Carlo (MC) simulation.
PYTHIA~8~\cite{pythia} explains the data within uncertainties. These results validate the
performance of the new ITS as well as the new reconstruction algorithms.

Figure~\ref{dngammadeta_mb} presents the $\mathrm{d}N_\mathrm{\gamma}/\mathrm{d}\eta$ (solid blue circles)
measured for MB events in p--Pb collisions at $\sqrt{s_{\rm NN}}$~=~5.02~TeV in the rapidity interval,
$2.3~<~\eta~<~3.9$ together with the measurements of charged-particle production (solid magenta circles)
at mid-rapidity~\cite{ChPrpPb}. The measurements are compared with the predictions from HIJING~\cite{HIJING}
and DPMJET~\cite{DPMJET} event generators. Both MC models considered here are in good agreement with
$\mathrm{d}N_\mathrm{ch}/\mathrm{d}\eta$ whereas $\mathrm{d}N_\mathrm{\gamma}/\mathrm{d}\eta$ is slightly
overpredicted by DPMJET. These results provide new constraints on model calculations to understand particle,
and in particular photon, production in p--Pb collisions.

\section{Summary}
\vspace{-0.2cm}
We have presented and compared the charged-particle pseudorapidity density measured in pp, p--Pb,
and Pb--Pb collisions at $\sqrt{s_{\rm NN}}$~=~5.02~TeV. A clear enhancement of particle production
at mid-rapidity in central Pb--Pb collisions compared to pp collisions is observed whereas a linear
scaling is seen for p--Pb collisions with respect to the pp collisions. By transforming the
pseudorapidity distributions to rapidity distributions we have found that the measurements for
all three collision systems follow a normal distribution in rapidity. We have analysed the pilot
beam data in pp collisions at $\sqrt{s}$ = 0.9~TeV using the new ITS detector and the brand new
computing framework and the performance of the whole setup has been studied, obtaining excellent
perspective in view of Run~3 data taking. Finally, we have also measured the
$\mathrm{d}N_\mathrm{\gamma}/\mathrm{d}\eta$ at forward rapidity which follows the trend of
charged-particle measurements at mid-rapidity.

\end{document}